\documentclass[aps,amsmath,amssymb,reprint,twocolumn,pra,superscriptaddress,notitlepage,showpacs,tightenlines]{revtex4-1}
\usepackage{graphicx}
\usepackage{dcolumn}
\usepackage{bm}
\usepackage{siunitx}
\usepackage{color}
\usepackage{amsmath}
\usepackage{float}
\usepackage{verbatim}
\usepackage{soul}
\usepackage[colorlinks,citecolor=blue,linkcolor=red,hyperindex,CJKbookmarks]{hyperref}
\usepackage[ruled,vlined]{algorithm2e}

\definecolor{pink}{rgb}{0.7922, 0.3373, 0.4118}

\begin{document}

\title{Full-quantum variational dynamics simulation for time-dependent Hamiltonians with global spectral discretization}

\author{Minchen Qiao}
\thanks{These authors contributed equally to this work.}

\author{Zi-Ming Li}
\thanks{These authors contributed equally to this work.}

\author{Yu-xi Liu}
\email{yuxiliu@mail.tsinghua.edu.cn}
\affiliation{School of Integrated Circuits, Tsinghua University, Beijing 100084, China}
\affiliation{Frontier Science Center for Quantum Information, Beijing, China}

\date{\today}


\begin{abstract}

The most widely used approach for simulating the dynamics of time-dependent Hamiltonians via quantum computation depends on the quantum-classical hybrid variational quantum time evolution algorithm, in which ordinary differential equations of the variational coefficients for determining time evolution are solved via classical simulations with a time discretization method.
We here present a full-quantum approach, in which ordinary differential equations of the variational coefficients are transformed into static linear equations via the Chebyshev spectral discretization method and then solved via the quantum singular value transformation algorithm.
Our full quantum algorithm avoids classical feedback, achieves exponential convergence for smooth Hamiltonians, and yields a quantum circuit depth that is independent of the number of time steps.
We demonstrate two implementation strategies, with a global formulation designed for fault-tolerant architectures and a sequential formulation tailored to near-term devices, and validate the approach through numerical simulations of proton--hydrogen charge-transfer dynamics, a prototypical time-dependent quantum chemistry problem.
This work establishes a systematic pathway from quantum-classical hybrid variational quantum algorithms to full-quantum solvers for general time-dependent Hamiltonians, particularly those whose dynamics admit compact variational descriptions, opening a route toward full quantum computational advantages in time-dependent simulations.

\end{abstract}
\maketitle

\section{Introduction}

Hamiltonian simulation constitutes one of the most natural and fundamental applications of quantum computation, providing a direct route to probing the dynamical behavior of quantum systems~\cite{lloyd1996universal, miessen2023quantum}.
While substantial algorithmic progress has been achieved for time-independent Hamiltonians~\cite{childs2018toward, low2019hamiltonian, childs2021theory}, many phenomena of central importance are governed by time-dependent interactions.
Such time-dependent Hamiltonian dynamics arise ubiquitously across atomic and molecular physics~\cite{eichler2005lectures, shi2024quantum}, quantum chemistry~\cite{helgaker2013molecular, maitra2016perspective, wahyutama2024simulating}, and quantum control~\cite{dong2010quantum, chen2013closed, khalid2023sample}.
A representative example is ion--atom charge transfer, where the electronic wavefunction evolves under a two-center Coulomb potential that varies continuously along the nuclear trajectory, posing a prototypical time-dependent quantum problem~\cite{fritsch1991semiclassical}.
In these scenarios, accurate simulation of quantum dynamics often requires resolving long-time evolution in high-dimensional Hilbert spaces as both system size and dynamical complexity increase.
As a result, time-dependent problems pose distinct algorithmic challenges beyond their time-independent counterparts~\cite{an2022time}.

A variety of quantum algorithms have been developed to address the dynamics simulation of time-dependent Hamiltonians, mainly containing quantum-classical hybrid variational and full-quantum approaches. The quantum-classical hybrid variational approaches, including quantum-assisted methods~\cite{bharti2021quantum, haug2022generalized, lau2022nisq} and quantum-parameterized methods~\cite{yuan2019theory, yao2021adaptive}, have been extensively explored in quantum chemistry and many-body physics, where parameterized quantum states are employed to reduce the effective dimension of the dynamics~\cite{mcardle2019variational}. Especially, it has recently been shown that the quantum-classical hybrid variational approach can be used to solve ion-atom collision dynamics~\cite{qiao2025quantum}. However, these quantum-classical hybrid algorithms rely highly on classical feedback, which limits their speed and performance.
Alternatively, there have been full-quantum solvers, including product formula methods~\cite{huyghebaert1990product, suzuki1993general, bosse2025efficient}, truncated Dyson series algorithms~\cite{low2018hamiltonian, kieferova2019simulating, berry2024quantum}, Magnus-based methods~\cite{blanes2009magnus, casares2024quantum}, and randomized schemes such as continuous qDRIFT~\cite{berry2020time}.
These methods do not require the assistance of classical computers. Instead, they directly approximate or stochastically sample the time-ordered evolution operator.
However, they operate at the operator level without dimensional reduction, while incurring circuit depths that grow with simulation time and the sparsity of the Hamiltonian.

We here develop a new full-quantum approach to solve the dynamics of the time-dependent Hamiltonian on the quantum computer by combining the dimensional reduction of variational methods with the coherent execution of fully quantum solvers, and bypassing both the deep-circuit bottleneck and the classical feedback overhead.
Our approach integrates variational parameterization, spectral discretization, and quantum linear system solvers for the dynamics simulation of time-dependent Hamiltonians.
The approach begins by reformulating the time-dependent Schr{\"o}dinger equation within a variational setting, yielding a set of coupled ordinary differential equations governing the evolution of variational parameters.
Next, by employing Chebyshev spectral expansions in time, these ordinary differential equations are mapped onto a time-independent linear system whose structure explicitly encodes the temporal dependence of the dynamics.
The resulting linear system is then solved using the quantum singular value transformation (QSVT) algorithm, allowing the quantum dynamics to be reconstructed without direct approximation of the time-ordered evolution operator.
Within this framework, both a global formulation and a sequential formulation emerge naturally as complementary realizations, adaptable to varying system sizes and quantum resource constraints.
We demonstrate our approach by using proton--hydrogen charge-transfer dynamics as an example.
This single-electron collision problem naturally satisfies the conditions under which variational subspace compression is most effective, as the physically accessible dynamics is confined to a low-dimensional subspace of the full Hilbert space, enabling substantial dimensional reduction with a compact ansatz.
The method applies in principle to any time-dependent Hamiltonian expressible in linear-combination-of-unitaries form.

The paper is organized as follows. In Sec.~\ref{sec:II}, we derive the variational projection of the time-dependent Schr\"odinger equation onto a low-dimensional subspace, discretize the equations of motion by using Chebyshev spectral methods, and construct both the global and sequential linear-system formulations. In Sec.~\ref{sec:III}, we show how to solve the resulting linear equations by using the QSVT-based quantum linear system algorithm. In Sec.~\ref{sec:IV}, we validate our approach by using proton--hydrogen charge-transfer dynamics as an example, examining spectral convergence, state fidelity, and quantum resource requirements for both formulations. Finally, conclusions and prospects for broader applications are presented in Sec.~\ref{sec:V}.


\section{Theoretical Framework}\label{sec:II}

In this section, we reformulate the time-dependent Schr{\"o}dinger equation by reducing it to a low-dimensional system of ordinary differential equations via variational parameterization, then discretizing the resulting dynamics through the spectral method, and finally encoding the problem as a linear system amenable to quantum computation with quantum linear solvers.

\subsection{Variational parameterization}

The dynamics of a closed quantum system governed by a time-dependent Hamiltonian $H(t)$ is described by the time-dependent Schr{\"o}dinger equation,
\begin{equation}
i \frac{d}{dt} |\Psi(t)\rangle = H(t) |\Psi(t)\rangle.
\label{eq:tdse}
\end{equation}
We consider the time-dependent Hamiltonians that can be expressed in the linear combination of unitaries (LCU) form:
\begin{equation}
    H(t)=\sum_{\gamma}g_{\gamma}(t)H_{\gamma},
\label{eq:LCU}
\end{equation}
where $g_\gamma(t)\in\mathbb{C}$ are time-dependent coefficients and $H_{\gamma}=\bigotimes_{p=0}^{N_Q-1}\sigma_{p}$ are $N_Q$-qubit Pauli strings, each defined as a tensor product of single-qubit Pauli operators.
This Pauli-sum representation arises naturally in quantum many-body systems, where second-quantized Hamiltonians are mapped onto qubits via standard fermion-to-qubit transformations such as the Jordan--Wigner or Bravyi--Kitaev mappings~\cite{lee2022simulating, tranter2018comparison}.
More broadly, any bounded operator on a finite-dimensional Hilbert space admits a decomposition into Pauli strings, making this framework applicable beyond the quantum many-body context~\cite{van2020circuit,mbeng2024quantum, schweizer2019floquet}.

To reduce the computational complexity of solving Eq.~(\ref{eq:tdse}), we project the dynamics onto a variational subspace.
Specifically, we represent the evolving quantum state as
\begin{equation}
|\Psi[\boldsymbol{\alpha}(t)]\rangle=\sum_{i=0}^{N_\alpha-1}\alpha_{i}(t)|\varphi_{i}\rangle,
\label{eq:ansatz}
\end{equation}
where $\boldsymbol{\alpha}(t) = [\alpha_0(t), \ldots, \alpha_i(t), \ldots, \alpha_{N_\alpha-1}(t)]^\top \in \mathbb{C}^{N_\alpha}$ denotes $N_\alpha$ time-dependent variational parameters corresponding to $N_\alpha$ variational quantum states $|\varphi_{i}\rangle$, and ${|\varphi_i\rangle}= U_i |\varphi_0\rangle$ with $U_0 =I$  are orthonormal basis states constructed by acting a set of unitaries $U_i$ on the reference state $|\varphi_0\rangle$.
In this work, we employ the simplified cumulative K-moment ansatz~\cite{qiao2025quantum}, constructed from the LCU Hamiltonian in Eq.~(\ref{eq:LCU}).
This ansatz yields a linearly independent and orthonormal basis that effectively captures the relevant dynamics of the system.

We next apply McLachlan's variational principle~\cite{mclachlan1964variational} to derive the equation of motion for $\boldsymbol{\alpha}(t)$, thereby minimizing the norm of the residual between the exact and variational time evolutions.
This yields the following evolution equation:
\begin{equation}
\frac{d\boldsymbol{\alpha}(t)}{dt}=\left[-i\sum_{\gamma}g_{\gamma}(t)\mathbf{N}^{-1}\mathbf{M}_{\gamma}\right]\boldsymbol{\alpha}(t),
\end{equation}
where the overlap matrix $\mathbf{N}$ and Hamiltonian coupling matrices  $\mathbf{M}_\gamma$ are defined as
\begin{equation}\label{eq:5}
\begin{aligned}
    N_{ij}&=\langle\varphi_{i}|\varphi_{j}\rangle,\\
    M_{\gamma,ij}&=\langle\varphi_{i}|H_{\gamma}|\varphi_{j}\rangle.
\end{aligned}
\end{equation}
These matrix elements can be computed classically if the ansatz is efficiently simulable, or estimated on a quantum computer using the direct measurement~\cite{qiao2025quantum, mitarai2019methodology}.
For convenience, we define the effective coefficient matrix
\begin{equation}
\mathbf{A}(t) \equiv -i\sum_{\gamma}g_{\gamma}(t)\mathbf{N}^{-1}\mathbf{M}_{\gamma} ,
\end{equation}
with $\mathbf{A}(t)\in \mathbb{C}^{N_\alpha \times N_\alpha}$, reducing Eq.~(\ref{eq:tdse}) to a system of $N_\alpha$ coupled linear ordinary differential equations:
\begin{equation}
\frac{d\boldsymbol{\alpha}(t)}{dt}=\mathbf{A}(t)\boldsymbol{\alpha}(t).
\label{eq:ode}
\end{equation}
The time dependence of $\mathbf{A}(t)$ is inherited entirely from the coefficients $g_\gamma(t)$ in the LCU decomposition.
When the coefficients $g_\gamma(t)$ are smooth, $\mathbf{A}(t)$ inherits this smoothness, which will be exploited by the spectral discretization introduced in the next subsection to achieve exponential convergence.
For Hamiltonians with discontinuous or rapidly varying coefficients, the evolution interval can be partitioned so that $\mathbf{A}(t)$ remains smooth within each segment.

The efficiency of the variational reduction critically depends on the dimension of the chosen ansatz space.
For systems with low excitation, such as single-electron problems or few-channel collision processes, the physically accessible subspace is typically much smaller than the full Hilbert space dimension $2^{N_Q}$, and the variational compression is correspondingly effective.
In more complex settings, such as strongly correlated or highly entangled many-body systems, $N_\alpha$ may grow rapidly, potentially diminishing the practical advantage of the variational projection.
The applicability of the present framework to such regimes remains an open question.
Here, we adopt the simplified cumulative K-moment ansatz for its systematic coverage of the dynamically reachable subspace, but other choices, such as natural orbital truncation based on occupation numbers or configuration-interaction states motivated by physical excitation channels, may also effectively depend on the specific systems~\cite{deprince2013accurate, hackl2020geometry, kossoski2023state}.

It is noted that previous variational principle based methods~\cite{qiao2025quantum, mitarai2019methodology, bharti2021quantum}  solve Eq.~(\ref{eq:ode}) to obtain variational parameters  $\boldsymbol{\alpha}(t)$ via classical computers after the matrix elements in Eq.~(\ref{eq:5}) are obtained via the quantum computation. We here propose an approach to solve Eq.~(\ref{eq:ode}) via quantum computation with spectral discretization and a quantum linear-system solver in the following subsections. This avoids the classical feedback and yields a full-quantum approach for simulating the dynamics of time-dependent Hamiltonians.

\subsection{Spectral discretization}

Having reduced the time-dependent Schr{\"o}dinger equation to a system of reduced linear ordinary differential equations, we now discretize the resulting dynamics using Chebyshev spectral methods~\cite{childs2020quantum}.
Spectral discretization is particularly well suited for problems with smooth time dependence, as it offers exponential convergence and enables efficient encoding of time evolution~\cite{trefethen2019approximation}.

To ensure numerical stability and control the spectral norm of the coefficient matrix, we partition the total evolution interval $[0,\,T]$ into $N_\tau$ subintervals:
\begin{equation}
[0,T] = [T_0,T_1] \cup [T_1,T_2] \cup \cdots \cup [T_{N_\tau-1},T_{N_\tau}],
\end{equation}
with $T_0=0$ and $T_{N_\tau}=T$.
Each subinterval $[T_h, T_{h+1}]$ is rescaled to the canonical Chebyshev domain $[-1,1]$ via the map:
\begin{equation}
K_h : t \mapsto t'=1-\frac{2(t-T_h)}{T_{h+1}-T_{h}},
\end{equation}
which satisfies $K_h(T_h) = 1$ and $K_h(T_{h+1}) = -1$.
Under this rescaling, we have $\boldsymbol{\alpha}_h(t') = \boldsymbol{\alpha}(t)$, and the evolution equation on the $h$-th subinterval becomes
\begin{equation}
    \frac{d\boldsymbol{\alpha}_h(t')}{dt'} = \mathbf{A}_h(t') \boldsymbol{\alpha}_h(t'),
\label{eq:ODE_h}
\end{equation}
with the coefficient matrix rescaled as
\begin{equation}\label{eq:11}
    \quad \mathbf{A}_h(t') \equiv \frac{T_{h} - T_{h+1}}{2} \cdot \mathbf{A}(t).
\end{equation}
Here, we enforce continuity at subinterval boundaries by setting $\boldsymbol{\alpha}_0(1) = \boldsymbol{\alpha}(0)$, and $\boldsymbol{\alpha}_{h+1}(1) = \boldsymbol{\alpha}_{h}(-1)$.

The choice of subinterval boundaries $\{T_h\}$ affects both computational cost and numerical stability.
For uniform segmentation, a sufficient condition to ensure $\lVert\mathbf{A}_h(t')\rVert \leq 1$ on each subinterval is given by~\cite{childs2021theory}:
\begin{equation}
\frac{T}{2N_\tau} \cdot \max_{t \in [0,T]} \lVert \mathbf{A}(t) \rVert \leq 1.
\label{eq:seg_equal}
\end{equation}
However, when $\lVert\mathbf{A}(t)\rVert$ varies significantly over time, uniform segmentation may lead to an unnecessarily large number of subintervals decided by the worst-case norm.
To improve efficiency, we adopt an adaptive segmentation strategy~\cite{an2022time}, which partitions the interval according to the cumulative norm:
\begin{equation}
\int_{T_h}^{T_{h+1}} \frac{1}{2}\lVert \mathbf{A}(t) \rVert dt = \frac{1}{2N_\tau} \int_0^T \lVert \mathbf{A}(t) \rVert dt \leq 1.
\label{eq:seg_adapt}
\end{equation}
This adaptive strategy concentrates computational resources by assigning shorter subintervals in regions where $\lVert\mathbf{A}(t)\rVert$ is large, and longer subintervals where it is small.
As a result, it often reduces the total number of required subintervals while maintaining accurate Chebyshev approximation with a moderate expansion degree.

Once the time domain is partitioned, we approximate the solution $\boldsymbol{\alpha}_h(t')$ on each subinterval using Chebyshev interpolation.
Specifically, each component ${\alpha}_{h,i}(t')$ of the vector $\boldsymbol{\alpha}_h(t')$  is expanded in a degree-$n$ Chebyshev basis:
\begin{equation}
{\alpha}_{h,i}(t') = \sum_{k=0}^{n} {c}_{h,i,k} T_k(t'),
\label{eq:Cheby}
\end{equation}
where
\begin{equation}
    \quad T_k(t') = \cos(k \arccos t'),
\end{equation}
are the Chebyshev polynomials of the first kind, and $\{c_{h,i,k}\}$ are the expansion coefficients.
To discretize the dynamics, we enforce Eq.~(\ref{eq:ODE_h}) at the  Chebyshev-Gauss-Lobatto collocation points $\{t'_l = \cos(\frac{l\pi}{n})\}_{l=0}^n$, resulting in a discrete system of linear equations that governs the coefficients $\{c_{h,\,i,\,k}\}$.
This collocation procedure yields the spectrally discretized form of the variational dynamics on each subinterval.
In the following subsection, we construct two formulations of this discretized linear system, including a global formulation and a sequential variant, and discuss their respective trade-offs for quantum implementation.

\subsection{Linear system constructions}

The collocation procedure described in the previous subsection yields a discrete system of linear equations on each time subinterval.
We now reorganize these equations into two complementary formulations tailored to different quantum resource regimes.
The first one is called as a global linear method~\cite{childs2020quantum}, which encodes the evolution across all subintervals into a single large-scale linear system via the quantum states. The second one is called as a sequential linear method, which is a generalization of the global one. We note that this generalization is non-trivial. In this case,  each subinterval is encoded in the quantum state independently, enabling more flexible resource allocation and modular quantum execution. The final post-processing with algebraic operations in the global formulation can be avoided.

\subsubsection{Global linear system}

To obtain the entire time evolution in a single quantum measurement, a global linear system, proposed in Ref.~\cite{childs2020quantum}, can be constructed via simultaneously encoding the spectral discretization across all $N_\tau$ subintervals.
This formulation incorporates both the collocation equations at Chebyshev-Gauss-Lobatto nodes and the continuity constraints linking adjacent temporal segments.
This yields a structured global linear system of the form~\cite{childs2020quantum}
\begin{equation}
    \mathbf{L}|X\rangle=|B\rangle
    \label{eq:global}
\end{equation}
where $|X\rangle \in \mathbb{C}^{N_\tau \cdot N_\alpha \cdot (n+1)}$ encodes the Chebyshev expansion coefficients $\{c_{h,\,i,\,k}\}$ across all time subintervals, and $|B\rangle$ embeds the initial condition.
Both vectors are represented as quantum states that can be naturally encoded across multiple registers, enabling the application of quantum linear system algorithms (QLSAs).

\begin{figure*}
    \centering
    \includegraphics[width=0.75\linewidth]{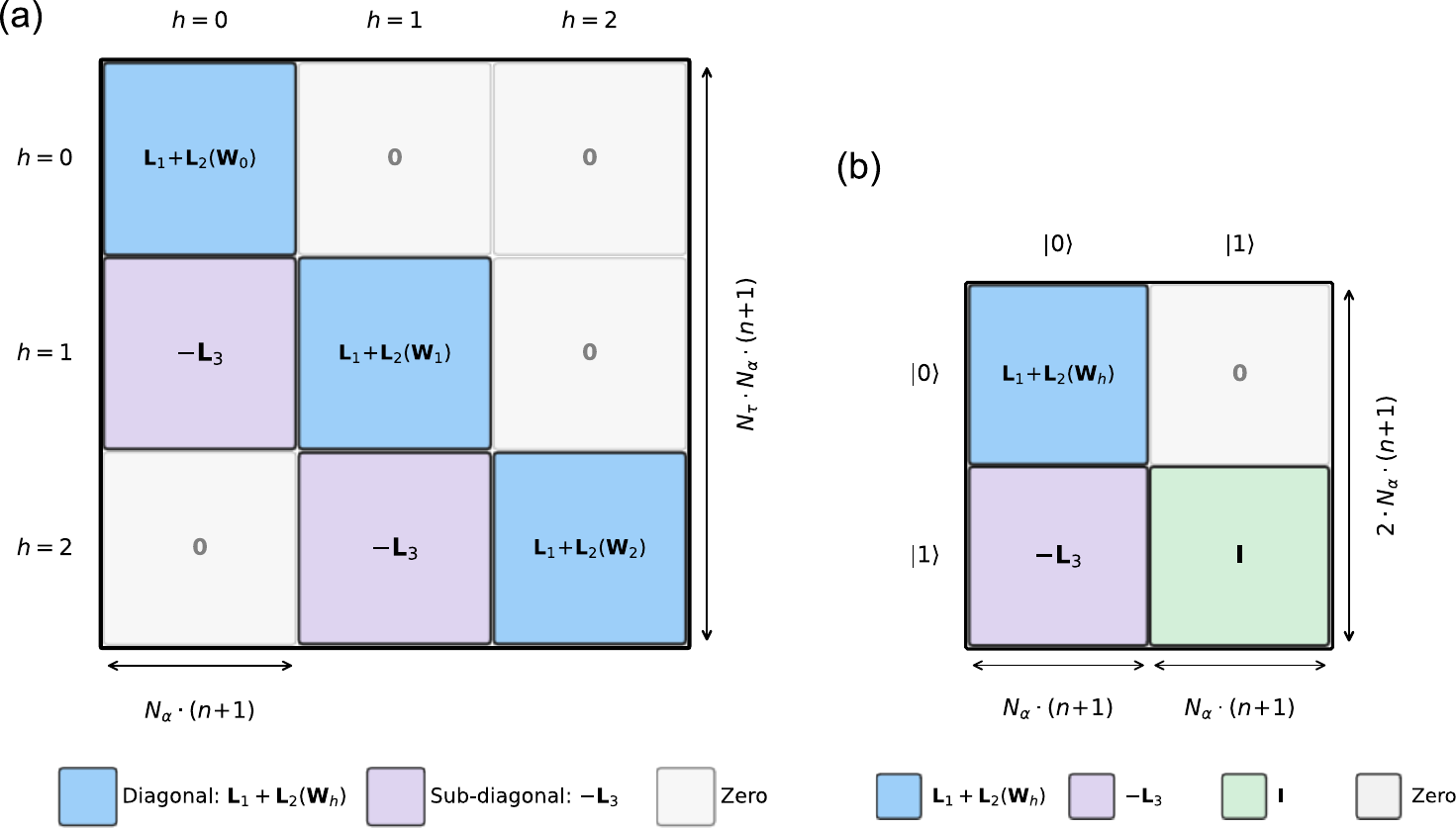}
    \caption{Block structure of the coefficient matrices in the two linear system formulations.
        (a) The global matrix $\mathbf{L}$ exhibits a block lower-triangular structure, illustrated here for $N_\tau = 3$ subintervals. Diagonal blocks $\mathbf{L}_1 + \mathbf{L}_2(\mathbf{A}_h)$ encode the collocation equations and initial conditions within each subinterval, while the sub-diagonal blocks $-\mathbf{L}_3$ enforce continuity between adjacent subintervals. Each block has dimension $N_\alpha(n+1)$.
        (b) The sequential matrix $\mathbf{L}_h$ for a single subinterval has a $2 \times 2$ block structure labeled by an index qubit. The upper-left block $\mathbf{L}_1 + \mathbf{L}_2(\mathbf{A}_h)$ implements the spectral collocation equations with dimension $N_\alpha(n+1)$. The lower-left block $-\mathbf{L}_3$ extracts the endpoint values from the Chebyshev coefficients. The lower-right block $\mathbf{I}$ preserves the extracted state with dimension $N_\alpha(n+1)$.}
    \label{fig:L_block}
\end{figure*}

We define the global solution vector of Eq.~(\ref{eq:global}) as
\begin{equation}
    |X\rangle=\sum_{h=0}^{N_\tau-1}\sum_{i=0}^{N_\alpha-1}\sum_{k=0}^n c_{h,i,k}|h\rangle|i\rangle|k\rangle
\end{equation}
where $c_{h,i,k}$ denotes the $k$-th Chebyshev coefficients for the component $i$ on the interval $h$.
$|h\rangle\otimes|i\rangle\otimes|k\rangle$ denotes states in three registers, which are used to encode the time interval, the components of the vector $\boldsymbol{\alpha}_h(t')$, and the order of the Chebyshev expansion. Correspondingly, we define $|{B}\rangle$ as
\begin{equation}
    |B\rangle=|0\rangle\left(\sum_{i=0}^{N_\alpha-1}\alpha_i(0)|i\rangle\right) |0\rangle
\end{equation}
with $\alpha_i(0)$ denoting the $i$-th component of the initial variational parameters $\boldsymbol{\alpha}(0)$, serving also as the boundary condition on the first subinterval. Here the left $|0\rangle$ denotes that the first register is in the ground state,  the right  $|0\rangle$ denotes that the third register is also in the ground state.
$\sum_{i=0}^{N_\alpha-1}\alpha_i(0)|i\rangle$ denotes the state of the second register and is used to encode the amplitudes of $N_\alpha$ variational parameters $\alpha_i(t)$ in Eq.~(\ref{eq:ansatz}) at the time $t=0$.

We now detail the structure of the matrix $\mathbf{L}$, which couples the dynamics of subintervals through both the local interpolation constraints and continuity conditions.
Specifically, we decompose $\mathbf{L}$ as
\begin{equation}
    \mathbf{L}=\sum_{h=0}^{N_\tau-1}|h\rangle\langle h|\otimes(\mathbf{L}_1+\mathbf{L}_2(\mathbf{A}_h))-\sum_{h=1}^{N_\tau-1}|h\rangle\langle h-1|\otimes \mathbf{L}_3.
    \label{eq:L_global}
\end{equation}
where the diagonal blocks $\mathbf{L}_1+\mathbf{L}_2(A_h)$ encode the collocation equations within each subinterval, and the sub-diagonal blocks $\mathbf{L}_3$ guarantee continuity between adjacent subintervals, as illustrated in Fig.~\ref{fig:L_block}(a).
The operator $\mathbf{L}_1$ is defined as:
\begin{equation}\label{eq:20}
    \mathbf{L}_1=\mathbf{I}_{N_\alpha}\otimes\left(|0\rangle\langle0|\mathbf{P}_n+\sum_{l=1}^n|l\rangle\langle l|\mathbf{P}_n\mathbf{D}_n \right)
\end{equation}
where $\mathbf{P}_n$ is the interpolation differentiation matrix with $[P_n]_{k,l} = \cos\frac{kl\pi}{n}$,
and $\mathbf{D}_n$ is the spectral differentiation matrix, whose nonzero elements are defined as
\begin{equation}
    [D_n]_{kl}=\frac{2l}{\sigma_k}, \quad k+l\mathrm{~odd},\,l>k
\end{equation}
with $\sigma_0=2$ and $\sigma_k=1$ for $k\in\{1,2,\cdots, n\}$.
Their product $\mathbf{P}_n\mathbf{D}_n$ maps the Chebyshev expansion coefficients $\{c_{h,i,k}\}$ to those of the time derivative, implementing spectral differentiation in coefficient space.
The second component, $\mathbf{L}_2(\mathbf{A}_h)$, incorporates the rescaled coefficient matrix evaluated at the Chebyshev nodes $t'_l$:
\begin{equation}\label{eq:22}
    \mathbf{L}_2(\mathbf{A}_h)=-\sum_{l=1}^n \mathbf{A}_h(t'_l)\otimes|l\rangle\langle l|\mathbf{P}_n,
\end{equation}
with $\mathbf{A}_h(t'_l)=\sum_{i_1,i_2=0}^{d-1}A_h(t'_l)_{i_1+1,i_2+1}|i_1\rangle\langle i_2|$, where $A_h(t'_l)_{i_1+1,i_2+1}$ denotes the element in the $(i_1 +1)$th row and the $(i_2+1)$th column of matrix $A_h(t'_l)$ as defined in Eq.~(\ref{eq:11}). Lastly, $L_3$ realizes the continuity condition between adjacent segments by transferring the endpoint information of the interval $h-1$ to the initial interval $h$:
\begin{equation}
    \mathbf{L}_3=\sum_{i=0}^{N_\alpha-1}\sum_{k=0}^n (-1)^{k}|i\rangle\langle i|\otimes|0\rangle\langle k|,
    \label{eq:L3}
\end{equation}
which implements $\alpha_{h,i}(-1) = \sum_{k=0}^{n}(-1)^k c_{h,i,k}$, exploiting the property $T_k(-1) = (-1)^k$.
The extracted state serves as the initial condition $\alpha_{h+1,i}(1)$ for the subsequent interval, ensuring continuity of the approximate solution across interval boundaries.

The resulting linear system offers a global encoding of the entire evolution, with an ideal structure that supports theoretical analysis and resource estimation.
However, its dimension grows rapidly with the number of subintervals and spectral resolution, making it more suitable for low-dimensional problems or as a conceptual foundation for quantum linear system algorithms.
Moreover, after finally solving the global linear equations via quantum computation,  a classical post-processing is still required to reconstruct the full time-dependent trajectory $\boldsymbol{\alpha}(t)$ in  Eq.~(\ref{eq:ansatz})  from the Chebyshev coefficients in Eq.~(\ref{eq:Cheby}) via simple algebraic operations.  In contrast, the quantum-classical hybrid variational quantum time evolution algorithm needs to classically solve differential equations, which is more difficult than the simple algebraic operation of the post-processing after solving the global linear equations via quantum computation. Below, we further generalize the global linear system~\cite{childs2020quantum} to the sequential linear system such that the classical post-processing can be avoided by enabling direct state preparation and measurement at each segment.

\subsubsection{Sequential linear system}

We now propose a sequential linear method to avoid the post-processing with algebraic operations in the global linear method by solving the time evolution interval by interval.
This method propagates the solution forward using only local information from the previous step.
Specifically, we define an independent linear system for each subinterval as
\begin{equation}
    \mathbf{L}_h|X_h\rangle=|B_h\rangle
    \label{eq:sequential}
\end{equation}
reducing the problem to $N_\tau$ decoupled systems, each of dimension $2 \times  N_\alpha(n+1)$ and thus significantly lowering the quantum resource requirements.
The matrix $\mathbf{L}_h$ for each interval admits the block-structured form
\begin{equation}
    \mathbf{L}_h = |0\rangle\langle 0| \otimes \left[ \mathbf{L}_1 + \mathbf{L}_2(\mathbf{A}_h) \right] - |1\rangle\langle 0| \otimes {\mathbf{L}}_3 + |1\rangle\langle 1|\otimes\mathbf{I},
    \label{eq:Lh_seq}
\end{equation}
with $\mathbf{L}_1$, $\mathbf{L}_2(\mathbf{A}_h)$, and $\mathbf{L}_3$ defined in Eqs.~(\ref{eq:20}), (\ref{eq:22}), (\ref{eq:L3}), respectively,
and $\mathbf{I}$ denotes the $N_\alpha(n+1)$-dimensional identity matrix.
Here the first register introduces an index qubit that partitions $\mathbf{L}_h$ into $2 \times 2$ blocks, as shown in Fig.~\ref{fig:L_block}(b), where the diagonal block $|0\rangle\langle 0|$ implements the spectral collocation equations, the off-diagonal block $|1\rangle\langle 0|$ extracts endpoint values via $\mathbf{L}_3$, and the diagonal block $|1\rangle\langle 1|$ preserves the extracted state.

The solution vector $|X_h\rangle$ comprises two components:
\begin{equation}
    |X_h\rangle=|0\rangle\sum_{i=0}^{N_\alpha-1}\sum_{k=0}^nc_{h,i,k}|i\rangle|k\rangle+|1\rangle\left(\sum_{i=0}^{N_\alpha-1}\alpha_{h,i}(-1)|i\rangle\right)|0\rangle,
\end{equation}
where the first term contains the Chebyshev expansion coefficients within subinterval $h$, and the second term stores the endpoint values at $t'=-1$ for each coefficient component.
The initial condition for subinterval $h$, represented by $|B_h\rangle$, takes the form
\begin{equation}
    |B_h\rangle=|0\rangle\left(\sum_{i=0}^{N_\alpha-1}\alpha_{h-1,i}(-1)|i\rangle\right)|0\rangle.
\end{equation}
This construction enables seamless propagation across interval boundaries.
By projecting the solution state $|X_h\rangle$ onto the index subspace $|1\rangle$ and postselecting, one directly extracts a normalized quantum state encoding $\boldsymbol{\alpha}_h(-1)$, which serves as the input $|B_{h+1}\rangle$ for the subsequent interval without additional classical processing.
The projection inherently enforces normalization at each subinterval, ensuring that the propagated state remains properly normalized throughout the sequential evolution.
If the postselection success probability is insufficient, amplitude amplification~\cite{brassard2000quantum}  can be employed to enhance it.

In summary, for both global and sequential linear methods, the output of the QLSA is encoded in the variational parameter space rather than the full Hilbert space.
The two methods differ in how the variational parameters are extracted.
The global one requires classical Chebyshev summation to reconstruct $\boldsymbol{\alpha}(t)$ from the expansion coefficients, whereas the sequential one directly provides the endpoint values $\boldsymbol{\alpha}_h(-1)$ in the $|1\rangle$ block of $|X_h\rangle$.
In either case, once $\boldsymbol{\alpha}(t)$ is obtained, the physical state is reconstructed via the ansatz as shown in Eq.~(\ref{eq:ansatz}).

The sequential formulation decomposes the global time evolution into a series of smaller, decoupled linear equations, each governing a single subinterval. By propagating the solution through local initial conditions, it eliminates the need for global coupling and classical post-processing.
This modular structure significantly reduces quantum circuit depth and memory requirements, making it more compatible with near-term quantum devices.


\section{Quantum Algorithm Implementation}\label{sec:III}

Quantum algorithms are applied to solve the linear system of equations as defined in Eq.~(\ref{eq:global}) and Eq.~(\ref{eq:sequential}). Currently, several quantum algorithms have been proposed for solving linear systems of equations, including methods based on the HHL algorithm~\cite{harrow2009quantum, wossnig2018quantum, childs2017quantum}, methods based on quantum linear algebra~\cite{li2025quantum}, and methods based on quantum signal processing~\cite{low2017optimal, gilyen2019quantum,martyn2021grand}. In our case, the matrix $\mathbf{L}$ or $\mathbf{L}_h$ as in Eq.~(\ref{eq:L_global}) and Eq.~(\ref{eq:Lh_seq}) are usually non-Hermitian matrices; thus we find that the method of quantum singular value transformation algorithm~\cite{gilyen2019quantum, martyn2021grand} suits the best for solving the linear equations which are constructed from the process of spectral discretization.

We first describe the quantum singular value transformation algorithm and its realization with quantum circuits.
For any operator $\mathbf{L}$ in the matrix representation with size $N_L$ acting on $n_L$ qubits,  its matrix representation has the singular value decomposition
\begin{equation}
	\mathbf{L}=\sum_{i=1}^{N_L} \sigma_i |v_i\rangle \langle w_i|,
\end{equation}
where each of the vector set $\{|v_i\rangle\}$ and $\{|w_i\rangle\}$ forms a complete set of vectors. The main purpose of quantum singular value transformation is to realize the operator $\mathcal{P}(\mathbf{L})$, which is a polynomial function of a matrix of order $d$, defined as
\begin{align}
	\mathcal{P}(\mathbf{L})&=\sum_{i=1}^{N_L} \mathcal{P}(\sigma_i) |v_i\rangle \langle w_i|\nonumber\\
	&=\sum_{i=1}^{N_L} \left(\sum_{j=0}^d a_j \sigma_i^j\right) |v_i\rangle \langle w_i|,
\end{align}
where the set of $\{a_0,a_1,\cdots,a_d\}$ is the set of  parameters that define the polynomial function $\mathcal{P}(\cdot)$.

\begin{figure}[!t]
	\includegraphics[width=\linewidth]{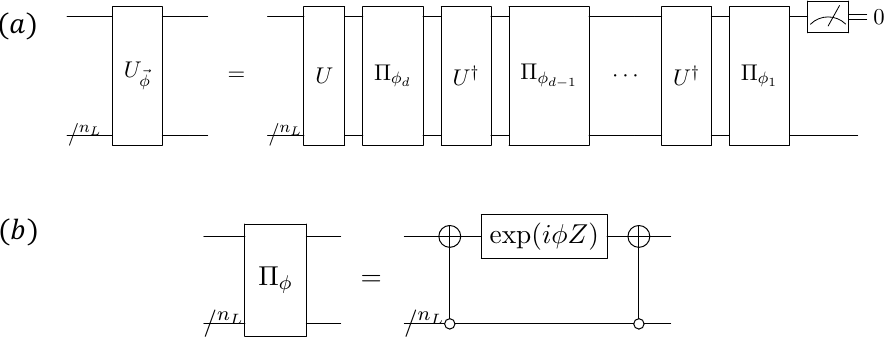}
	\caption{The realization of the quantum singular value transformation algorithm with quantum circuits. (a) The realization of operator $U_{\boldsymbol{\phi}}$, suppose the order $d$ is an even number. (b) The realization of each projector-controlled phase-shift operator $\Pi_{\phi}$.}
	\label{fig1}
\end{figure}

To realize the operator $\mathcal{P}(\mathbf{L})$, two other operators are defined on a composite system, containing $n_L$ system qubits and $1$ ancillary qubit, as shown in Fig.~\ref{fig1}. First, an operator $U$ is defined as
\begin{equation}
	U=\left(
	\begin{array}{cc}
		\mathbf{L} & \sqrt{1-\mathbf{L}\mathbf{L}^{\dagger}}\\
		\sqrt{1-\mathbf{L}^{\dagger}\mathbf{L}} & -\mathbf{L}^{\dagger}
	\end{array}
	\right),
\end{equation}
which is a unitary matrix that serves as a block encoding of the matrix $\mathbf{L}$. The block $\sqrt{1-\mathbf{L}\mathbf{L}^{\dagger}}$ is defined as
\begin{equation}
	\sqrt{1-\mathbf{L}\mathbf{L}^{\dagger}}=\sum_{i=1}^{N_L} \sqrt{1-\sigma_i^2} |v_i\rangle \langle v_i|,
\end{equation}
and the block $\sqrt{1-\mathbf{L}^{\dagger}\mathbf{L}}$ is defined as
\begin{equation}
	\sqrt{1-\mathbf{L}^{\dagger}\mathbf{L}}=\sum_{i=1}^{N_L} \sqrt{1-\sigma_i^2} |w_i\rangle \langle w_i|.
\end{equation}
It should be noted that this definition of $U$ requires all singular values $\{\sigma_i\}$ of the matrix $\mathbf{L}$ to be less than or equal to $1$. Otherwise, the matrix $\mathbf{L}$ should be normalized with a normalization factor $\alpha$, so that the maximum singular value of $\mathbf{L}/\alpha$ is less than or equal to $1$. In this case, the operator $U$ should be defined as
\begin{equation}
	U=\left(
	\begin{array}{cc}
		\mathbf{L}/\alpha & \sqrt{1-\mathbf{L}\mathbf{L}^{\dagger}/\alpha^2}\\
		\sqrt{1-\mathbf{L}^{\dagger}\mathbf{L}/\alpha^2} & -\mathbf{L}^{\dagger}/\alpha
	\end{array}
	\right),
\end{equation}
ensuring that the operator $U$ is still a unitary matrix.

Next, a projector-controlled phase-shift operator $\Pi_{\phi}$ is defined as
\begin{equation}
	\Pi_{\phi}=I\otimes \exp\left(i\phi \left(2|0\rangle\langle 0| -I\right)\right),
\end{equation}
which is a rotation operator around the $z$-axis acting on the system qubits. Given the operator $U$ and $\Pi_{\phi}$, for a sequence of phase factors $\boldsymbol{\phi}=\left(\phi_1,\phi_2,\cdots,\phi_d\right)$, an operator $U_{\boldsymbol{\phi}}$ is defined as
\begin{equation}
	U_{\boldsymbol{\phi}}=\Pi_{\phi_1}U\Pi_{\phi_2}U^{\dagger}\cdots \Pi_{\phi_{d-1}}U^{\dagger}\Pi_{\phi_d}U
\end{equation}
for odd $d$, and
\begin{equation}
	U_{\boldsymbol{\phi}}=\Pi_{\phi_1}U^{\dagger}\Pi_{\phi_2}U\cdots \Pi_{\phi_{d-1}}U^{\dagger}\Pi_{\phi_d}U
\end{equation}
for even $d$. By choosing the sequence of phase factors $\boldsymbol{\phi}$ appropriately based on the set of $\{a_0,a_1,\cdots,a_d\}$, which are the parameters that define the polynomial function $\mathcal{P}(\cdot)$, the operator  $U_{\boldsymbol{\phi}}$ is a block encoding of the operator $\mathcal{P}(\mathbf{L})$ as
\begin{equation}
		U_{\boldsymbol{\phi}}=\left(
	\begin{array}{cc}
		\mathcal{P}(\mathbf{L}) & \cdot\\
		\cdot & \cdot
	\end{array}
	\right),
\end{equation}
and the operator
\begin{equation}
	\langle 0| U_{\boldsymbol{\phi}} |0\rangle=\mathcal{P}(\mathbf{L})
\end{equation}
is a realization of  the operator $\mathcal{P}(\mathbf{L})$. The construction of the phase factor sequence $\boldsymbol{\phi}$ based on the polynomial function $\mathcal{P}(\cdot)$ can be done via mathematical approaches~\cite{gilyen2019quantum} or numerical optimization~\cite{haah2019product, dong2021efficient}.

The realization of the process of quantum singular value transformation with quantum circuits is shown in Fig.~\ref{fig1}.
The quantum circuits contain $n_L+1$ qubits. The top qubit is an ancillary qubit initialized in state $|0\rangle$, which helps the realization of each $U$ and $\Pi_{\phi}$.
The bottom $n_L$ qubits are the system qubits, initialized in state $|\psi_{\rm in}\rangle$.
The circuit is run and the ancillary qubit is measured at the end of the quantum circuit.
If the measurement outcome is $0$, then the whole quantum singular value transformation algorithm is successful. In this case, the output state $|\psi_{\rm out}\rangle$ of the system qubits is
\begin{equation}
	\label{qsvt}
	|\psi_{\rm out}\rangle=\frac{\mathcal{P}(\mathbf{L})|\psi_{\rm in}\rangle}{\parallel\mathcal{P}(\mathbf{L})|\psi_{\rm in}\rangle\parallel},
\end{equation}
indicating that the operator of $\mathcal{P}(\mathbf{L})$ has been realized.

We then explain how the method of quantum singular value transformation can be used for solving the linear system of equations. To solve a linear system of equation $\mathbf{L}|X\rangle=|B\rangle$ as shown in Eq.~(\ref{eq:global}), it is required that the operator $\mathbf{L}^{-1}$ is realized so that we have $|X\rangle=\mathbf{L}^{-1}|B\rangle$. If a polynomial function $\mathcal{P}$ with order $d$ is able to approximate the function of $f(x)=1/x$, as
\begin{equation}
	|\mathcal{P}(x)-\frac{1}{x}|=|\sum_{j=0}^d a_j x^j-\frac{1}{x}|<\frac{\epsilon}{N}, \forall \sigma_{min} \leq x\leq \sigma_{max},
\end{equation}
where $\sigma_{min}$ and $\sigma_{max}$ denotes the minimum and the maximum singular value of matrix $\mathbf{L}$, then we have
\begin{equation}
	\parallel \mathcal{P}(\mathbf{L})-f(\mathbf{L})\parallel\leq \parallel 	\sum_{i=1}^{N}|\sum_{j=0}^d a_j \sigma_i^j-\frac{1}{\sigma_i}| |v_i\rangle\langle w_i| \parallel<\epsilon.
\end{equation}
Thus, the polynomial function of the matrix $\mathcal{P}(\mathbf{L})$ gives a good approximation of the operator $f(\mathbf{L})=1/\mathbf{L}$. Given the polynomial function $\mathcal{P}(\cdot)$, the quantum singular value transformation algorithm is then applied to realize the operator $\mathcal{P}(\mathbf{L})$ with the input state of the system register taken as $|B\rangle$. If the process of the quantum singular value transformation algorithm is successful, then according to Eq.~(\ref{qsvt}), the output state $|\psi_{\rm out}\rangle$ satisfies
\begin{equation}
	|\psi_{\rm out}\rangle\propto \mathcal{P}(\mathbf{L})|B\rangle\approx \mathbf{L}^{-1}|B\rangle=|X\rangle.
\end{equation}
Therefore, the solution of the linear system of equations has been encoded as a quantum state, which is stored in the system qubits. The task of solving a linear system of equations is completed with the quantum singular value transformation.


\section{Numerical results}\label{sec:IV}

The proposed method is validated by simulating charge-transfer dynamics in proton--hydrogen collisions, a prototypical time-dependent quantum chemistry problem.
This system is well-suited as a benchmark because the Hamiltonian varies smoothly yet nontrivially along the nuclear trajectory, while the dynamics is governed by a single active electron and can be effectively captured within a compact subspace of the full Hilbert space.
We first specify the physical model and its quantum computational representation, then examine spectral convergence under the global formulation, and finally verify consistency with the sequential formulation and explicit quantum-circuit simulation.

\begin{figure}
	\centering
	\includegraphics[width=0.96\linewidth]{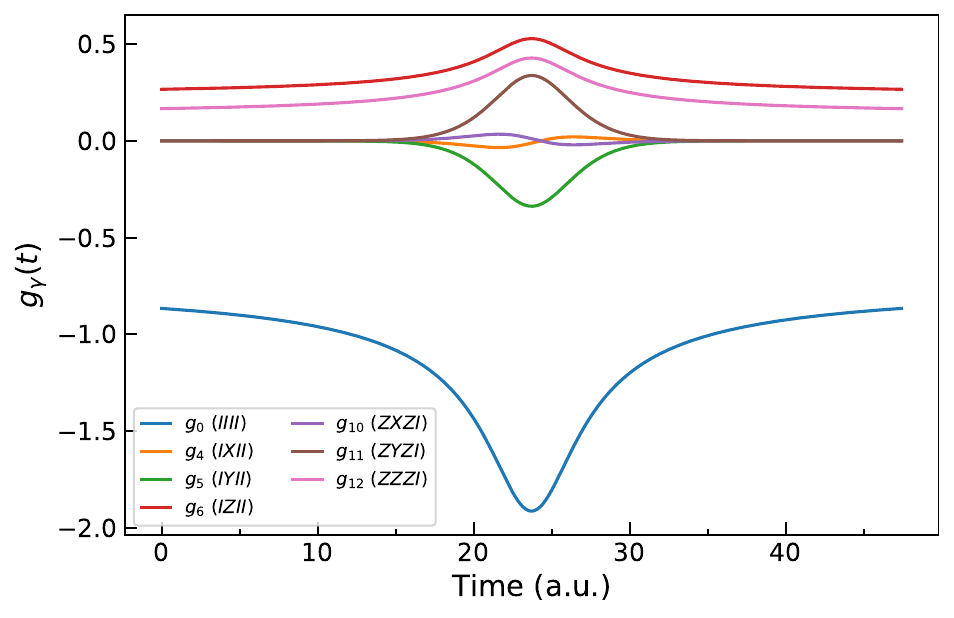}
	\caption{Representative time-dependent coefficients $\{g_\gamma(t)\}$ of the qubit Hamiltonian for the $\mathrm{H^+}+\mathrm{H}(1s)$ collision at $E=10$~keV and $b=1.6$~a.u.
    Seven distinct coefficients are shown, while the remaining terms exhibit identical temporal behavior and are omitted for clarity. }
	\label{fig:gt}
\end{figure}

\begin{figure*}
    \centering
    \includegraphics[width=0.9\linewidth]{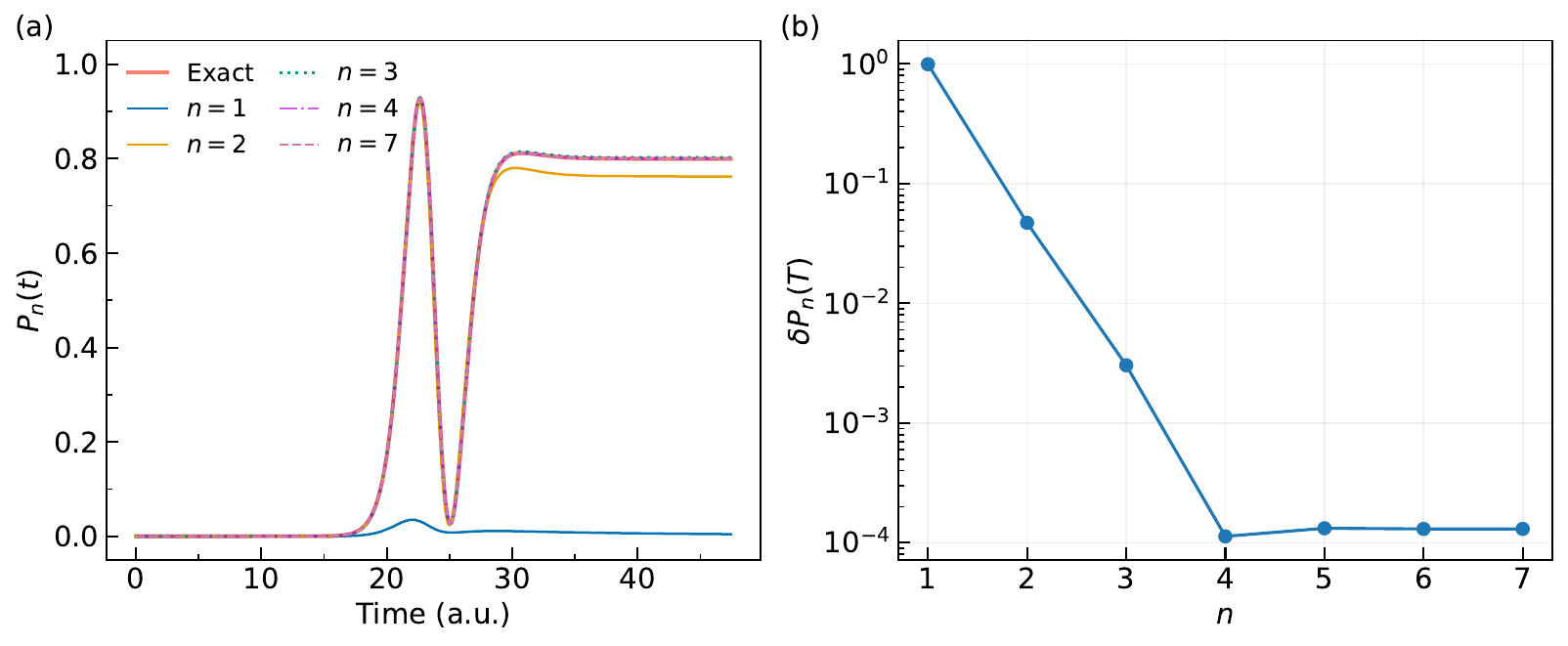}
    \caption{Spectral convergence of the global formulation for the $\mathrm{H^+}+\mathrm{H}$ charge-transfer dynamics.
    (a) Time-dependent charge-transfer probability $P_n(t)$ for Chebyshev degrees $n=1,2,3,4,7$, compared with the exact solution.
    Low-order expansions ($n=1,2$) fail to capture the asymptotic transfer probability, while $n\geq 3$ accurately reproduces both the transient dynamics and the long-time limit.
    (b) Relative error $\delta P_n(T)$ of the asymptotic charge-transfer probability as a function of the Chebyshev degree $n$. The error decreases exponentially up to $n=4$, beyond which it saturates at $\sim 10^{-4}$.}
    \label{fig:PPn}
\end{figure*}

We consider the resonant charge-transfer process $\mathrm{H}^+ + \mathrm{H}(1s) \rightarrow  \mathrm{H}(1s) + \mathrm{H}^+$ at collision energy $E = 10$~keV and impact parameter $b = 1.6$~a.u., following the setup of Ref.~\cite{qiao2025quantum}.
The projectile proton travels along a classical straight-line trajectory $\bm{R}(t) = \bm{b} + \bm{v}t$, and the electron evolves quantum mechanically under the resulting time-dependent two-center Coulomb potential.
At this collision energy, a two-state approximation is sufficient, and the spatial orbital space is truncated to the $1s$ states centered on each nucleus~\cite{mccarroll1961resonance}.
The electronic structure is described using the STO-3G (Slater-type Orbital-3 Gaussians) basis augmented with electron translation factors to incorporate the nonadiabatic motion of the atomic orbitals~\cite{bates1958electron}.
This construction yields two spatial orbitals, which, upon spin-orbital expansion, give rise to four spin orbitals.
The corresponding time-dependent qubit Hamiltonian takes the LCU form of Eq.~(\ref{eq:LCU}) with $N_Q=4$ qubits and $N_\gamma=13$ Pauli terms~\cite{qiao2025quantum}:
\begin{align}
	H(t) = & \, g_0(t)+  g_1(t)X_0  + g_2(t)Y_0 + g_3(t)Z_0 + g_4(t)X_2 \nonumber \\
	+ &\,  g_5(t)Y_2 + g_6(t)Z_2 +g_7(t)Z_1X_0 + g_8(t)Z_1Y_0 \nonumber \\
	+& \,   g_9(t)Z_1Z_0+g_{10}(t)Z_3X_2Z_1 + g_{11}(t)Z_3Y_2Z_1 \nonumber \\
	+&\, g_{12}(t)Z_3Z_2Z_1,
	\label{eq:Ht}
\end{align}
where the coefficients $\{g_\gamma(t)\}$ inherit the smooth time dependence of the original coordinate-space Hamiltonian after second quantization and Bravyi--Kitaev transformation, with the latter serving as a standard fermion-to-qubit encoding tool~\cite{bravyi2002fermionic}.
Representative coefficient dynamics are shown in Fig.~\ref{fig:gt}.
Since $[H(t_1),\,H(t_2)] \neq 0$ for $t_1 \neq t_2$ in general, the evolution cannot be reduced to a simple exponential of the time-integrated Hamiltonian and requires explicit time ordering, providing a nontrivial test of the spectral discretization scheme.

We first perform a variational preprocessing step to construct a compact dynamical subspace.
The variational subspace is constructed using the simplified cumulative $K$-moment ansatz.
For the present single active electron system, this procedure produces $N_\alpha = 4$ variational basis states, reducing the effective dynamical dimension from the full Hilbert space dimension $2^{N_Q}=16$ to a subspace of size $N_Q=4$, as described in Ref.~\cite{qiao2025quantum}.
This substantial compression ratio reflects the low excitation complexity inherent in single-particle collision dynamics, where only a few electronic channels participate in the charge-transfer process.

The above construction defines a compact yet nontrivial dynamical problem that retains the essential time dependence of the collision process.
In the following, the proton--hydrogen collision is used to assess the performance of the global formulation, focusing first on convergence and accuracy within the ideal linear-system inversion limit.
Subsequently, the sequential formulation is employed to analyze the feasibility and resource requirements of explicit quantum-circuit implementations.

\subsection{Numerical accuracy of the variational spectral discretization}

We first assess our variational spectral discretization framework using the global formulation under an ideal QSVT.
Throughout this subsection, Eq.~(\ref{eq:seg_equal}) is satisfied and a time register of $7$ qubits is used to encode $N_\tau = 2^7 = 128$ uniform subintervals over the total evolution interval $[0,\,T]$.
With $N_\tau^{\mathrm{global}}=128$ and $N_\alpha=4$ fixed, the Chebyshev degree is varied from $n=1$ to $7$ to assess spectral convergence, yielding a global linear system of dimension $N_\tau^{\mathrm{global}} \times N_\alpha \times  (n+1)$.

The time-dependent charge-transfer probability is defined as
\begin{equation}
	P_n(t) = \sum_{\sigma=\uparrow,\downarrow}
	\left|
	\langle \Psi_{1s_{\mathrm{P}},\sigma} \mid \Psi_n(t) \rangle
	\right|^2,
	\label{eq:Pt}
\end{equation}
where $|\Psi_{1s_{\mathrm{P}},\sigma}\rangle$ denotes the traveling $1s$ spin orbital with spin $\sigma$ centered on the projectile proton, and $|\Psi_n(t)\rangle$ is the reconstructed quantum state obtained from the global solution of $\mathbf{L}|X\rangle=|B\rangle$ with $n$ degree Chebyshev fitting.
Specifically, the Chebyshev coefficients $\{c_{h,\,i,\,k}\}$ are first obtained from the global linear system, from which the time-dependent variational parameters $\alpha_i(t)$ are recovered and subsequently mapped back to the physical quantum state $|\Psi_n(t)\rangle$.
As a reference, the exact quantum dynamics is computed using the QuTiP~\cite{johansson2012qutip}.

Figure~\ref{fig:PPn}(a) shows the time-dependent charge-transfer probability results $P_n(t)$ for representative values, e.g.,  $n=\{1,2,3,4,7\}$, compared with the exact solution.
At $n=1$, the spectral expansion fails to resolve the collision dynamics, yielding a qualitatively incorrect evolution and a severely underestimated asymptotic probability.
For $n=2$, the overall trend is captured, but the asymptotic charge-transfer probability remains substantially inaccurate.
For $n\geq 3$, both the transient oscillatory behavior near closest approach and the asymptotic limit are accurately reproduced.
The curves for $n=4$ and $n=7$ are visually indistinguishable from the exact result over the entire evolution interval, and also agree with the result of classical-quantum hybrid variational quantum time evolution algorithm~\cite{qiao2025quantum}.

The convergence of the spectral discretization is quantified in Fig.~\ref{fig:PPn}(b).
It plots the relative asymptotic error
\begin{equation}
\delta P_n(T) =
\frac{|P_n(T)-P_{\mathrm{exact}}(T)|}{P_{\mathrm{exact}}(T)}
\end{equation}
as a function of the Chebyshev degree $n$, where $T$ denotes the final simulation time at which the charge-transfer probability has effectively converged.
The error decreases approximately exponentially from $\mathcal{O}(1)$ at $n=1$ to $\sim 10^{-4}$ at $n=4$, consistent with the expected spectral convergence for smooth dynamical trajectories~\cite{trefethen2019approximation}.
For $n\geq 4$, the error saturates at $\sim 10^{-4}$, suggesting that further increasing the Chebyshev degree no longer improves the solution within the present reduced dynamical representation.

\begin{figure*}
    \centering
    \includegraphics[width=0.9\linewidth]{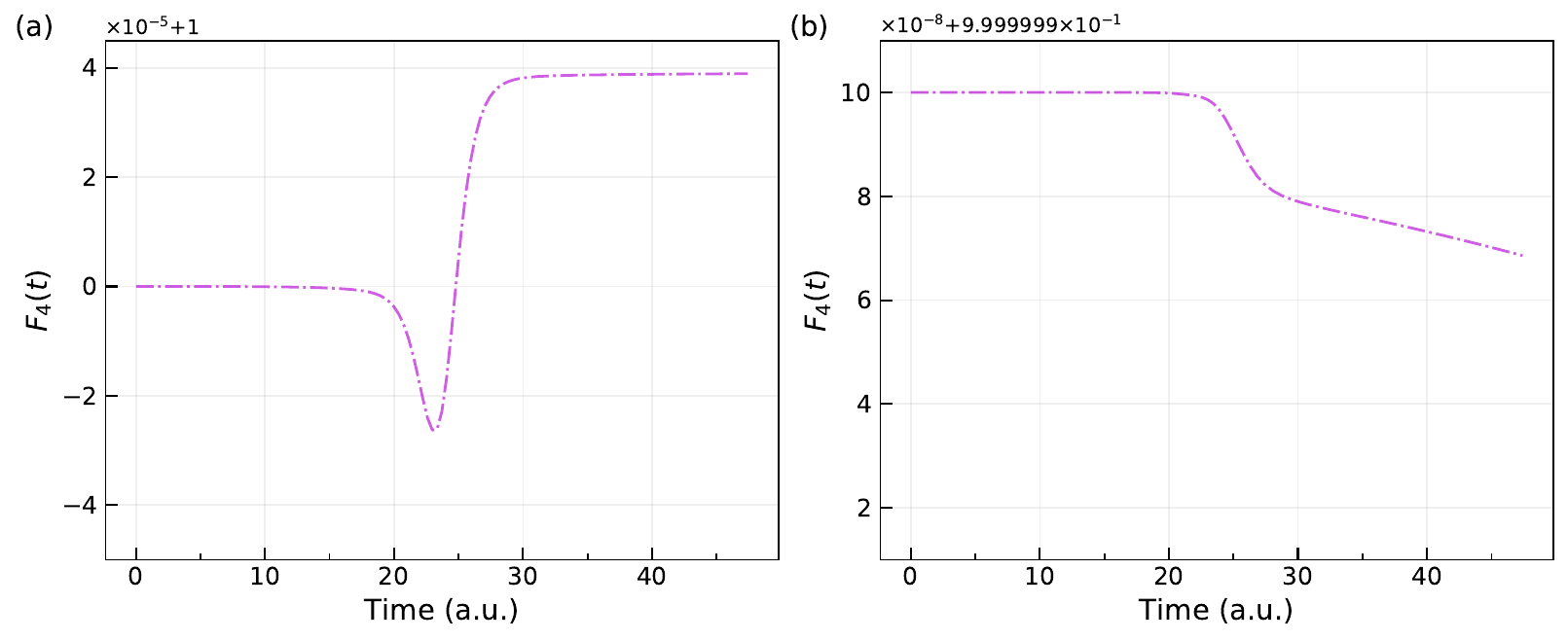}
    \caption{State fidelity for the $\mathrm{H^+}+\mathrm{H}$ collision with Chebyshev degree $n=4$.
    (a) Global formulation with equal segmentation.
    The fidelity of the global scheme exhibits deviations at the $10^{-5}$ level and reflects a norm drift level due to the absence of explicit intermediate normalization.
    (b) Sequential formulation with adaptive segmentation.
    The sequential scheme suppresses norm drift by normalizing at each step and maintains fidelity deviations at the level of $10^{-8}$.}
    \label{fig:F4}
\end{figure*}

To assess the accuracy across the entire evolution rather than only at the final time, Fig.~\ref{fig:F4}(a) presents the state fidelity
\begin{equation}
F_n(t)=|\langle\Psi_{\mathrm{exact}}(t)\mid\Psi_n(t)\rangle|^2
\end{equation}
obtained with Chebyshev degree $n=4$.
The fidelity remains above $99.997\%$ throughout the evolution, with a transient minimum near closest approach, where the Hamiltonian varies most rapidly.
The fidelity deviates slightly from unity after the collision region due to minor norm drift in the global formulation.
This deviation is purely numerical and arises from the absence of explicit intermediate normalization in the global linear-system construction.
It can be systematically suppressed by increasing the Chebyshev degree.
For instance, increasing the Chebyshev degree to $n=7$ reduces both fidelity and norm deviations to the $10^{-8}$ level throughout the evolution.
Alternatively, even at fixed $n=4$, a posteriori renormalization of the global state similarly suppresses the fidelity deviation to the $10^{-8}$ level, indicating that the residual discrepancy at $n=4$ arises primarily from norm drift rather than spectral truncation.
These results demonstrate exponential spectral convergence and confirm that the global variational spectral discretization accurately captures both transient and asymptotic features of the time-dependent collision dynamics.

\subsection{Sequential formulation and resource analysis}

We next examine the sequential formulation to verify its consistency with the global formulation and to assess the quantum resources involved in explicit circuit implementation.
Unlike the global results above, the sequential formulation is evaluated here through an explicit noiseless QSVT circuit implementation.
The same collision problem is solved with the Chebyshev degree fixed at $n=4$, for which spectral convergence has been established.
In the sequential formulation, adaptive time segmentation based on Eq.~(\ref{eq:seg_adapt}) is employed, reducing the number of subintervals from $128$  to $N_\tau^{\mathrm{seq}}=61$.
The adaptive segmentation halves the number of required QLSA invocations in the sequential formulation, directly reducing the total measurement cost.

As shown in Fig.~\ref{fig:F4}(b), the state fidelity between the sequential and exact solutions remains within numerical precision of unity throughout the evolution, confirming that the two formulations encode equivalent dynamics despite their distinct linear-system structures.
Beyond reproducing the global state evolution, the sequential formulation exhibits enhanced numerical stability.
This improvement arises from the stepwise normalization inherent in the sequential scheme, which suppresses the norm drift observed in the global formulation.
For $n=4$, the sequential fidelity deviation remains at $O(10^{-8})$ throughout the evolution, whereas the global solution exhibits deviations of order $O(10^{-5})$ near closest approach.
The explicit circuit simulations employ a QSVT polynomial of degree $d=1501$ to approximate $\mathbf{L}_h^{-1}$ at near-machine precision.
This corresponds to a QSVT circuit depth of approximately $6006$ layers, accounting solely for the polynomial transformation and excluding additional overhead from block-encoding decomposition and state preparation.
The chosen degree $d$ therefore represents a high-accuracy stress test of the solver rather than a practically optimized configuration.
Consistent with the theoretical scaling $d=\mathcal{O}(\kappa\log(\kappa/\epsilon))$, the required degree decreases substantially when moderate precision $\epsilon$ is sufficient~\cite{martyn2021grand}.

We now compare the quantum resource requirements of the global and sequential formulations.
In the global formulation, all $N_\tau$ subintervals are coupled into a single linear system of dimension $N_L^{\mathrm{global}} = N_\tau N_\alpha (n+1)$.
The corresponding quantum register requires  $\lceil \log_2 N_L^{\mathrm{global}} \rceil$ system qubits plus one ancillary qubit for QSVT.
In contrast, the sequential formulation solves $N_\tau$ independent systems, each of fixed dimension $ N_L^{\mathrm{seq}} = 2N_\alpha (n+1)$,  independent of $N_\tau$.
This requires $\lceil \log_2 N_L^{\mathrm{seq}} \rceil + 1$ qubits in total, where $\lceil \log_2 N_L^{\mathrm{seq}} \rceil$ qubits encode the linear system including the index qubit defined in Eq.~(\ref{eq:Lh_seq}), and one additional qubit serves as the QSVT ancilla.
The total qubit count is therefore reduced by approximately $\lceil \log_2 N_\tau \rceil$ compared with the global construction.

In the sequential formulation, the matrices $\mathbf{L}_h$ are confined to single subintervals, and their condition numbers remain bounded.
By contrast, the global matrix $\mathbf{L}$ couples all subintervals into a single system, and its condition number generally increases with system size.
Since the QSVT circuit depth also scales as $\mathcal{O}(\kappa \log(\kappa/\epsilon))$, the bounded $\kappa_{\mathrm{seq}}$ ensures that each sequential step can be implemented with moderate circuit depth, whereas the global formulation may require substantially deeper circuits.
In particular, the circuit depth depends on the condition number rather than directly on the number of subintervals $N_\tau$.

The two formulations also differ in how the dynamical information is distributed across QLSA invocations.
In the global formulation, the entire trajectory is encoded in a single quantum state and obtained from one QLSA invocation followed by measurement.
The sequential formulation, by contrast, produces only the endpoint value at each subinterval and therefore requires $N_\tau$ independent QLSA calls and measurements.
Accordingly, while the circuit depth of each sequential step remains moderate due to bounded $\kappa_{\mathrm{seq}}$, the total computational cost scales linearly with $N_\tau$.
The key structural distinctions between the two formulations are summarized in Table~\ref{tab:resource}.

\begin{table}[b]
    \caption{Summary of key structural differences between the global and sequential formulations.}
	\label{tab:resource}
        \renewcommand{\arraystretch}{1.5}
	\begin{ruledtabular}
		\begin{tabular}{lcc}
			& Global & Sequential \\
			\hline
			Linear-system dimension & $N_{\tau}^{\mathrm{global}} \cdot N_\alpha(n{+}1)$ & $2N_\alpha(n{+}1)$ \\
			QLSA invocations & 1 & $N_\tau^{\mathrm{seq}}$ \\
            Condition-number scaling  & $\kappa_{\mathrm{global}}$ & $\kappa_{\mathrm{seq}}$ \\
			Output & Full trajectory & Single-step \\
		\end{tabular}
	\end{ruledtabular}
\end{table}

To provide a concrete illustration, we now instantiate the above general analysis for the present collision problem.
In this case, the variational dimension is $N_\alpha=4$, the Chebyshev degree is $n=4$, and uniform segmentation uses $N_\tau^{\mathrm{global}}=128$ subintervals, while adaptive segmentation reduces this to $N_\tau^{\mathrm{seq}}=61$.
Under uniform segmentation, the global linear system has dimension $2560$, requiring $13$ qubits including the QSVT ancilla, and is solved with a single QLSA invocation.
In the sequential formulation, each subinterval corresponds to a system of dimension $40$, requiring $7$ qubits, and adaptive segmentation reduces the total number of QLSA calls to $61$.
For this problem, $\kappa_{\mathrm{seq}} \approx 40$ remains well controlled, while $\kappa_{\mathrm{global}}$ is significantly larger, though its precise value is problem dependent.

Overall, the two formulations exhibit complementary advantages.
For near-term implementations where circuit depth is the primary constraint, the sequential formulation with adaptive segmentation offers a practical path forward.
For fault-tolerant architectures capable of supporting deeper circuits, the global formulation fully exploits the advantage of obtaining the complete time evolution from a single quantum solve.


\section{Discussion and Conclusions}\label{sec:V}

In this work, we have introduced a fully quantum method to solve the time-dependent Hamiltonian dynamics by combining variational subspace compression, global Chebyshev spectral discretization, and quantum singular value transformation.
The variationally projected Schr{\"o}dinger equation is reformulated as a structured static linear system and solved via the QSVT algorithm, thereby avoiding explicit construction of time-ordered propagators and eliminating quantum-classical feedback loops inherent in hybrid variational methods.
Both the global and sequential constructions operate within this equation-level formulation, differing only in the organization of the spectral system.
The global realization defines the conceptual core by encoding the full evolution into a single coherent linear problem, while the sequential variant reorganizes this structure to moderate conditioning effects in long-time simulations.
Numerical benchmarks for proton--hydrogen charge-transfer dynamics confirm the internal consistency and high-precision performance of the framework.

Conceptually, the present formulation shifts the focus of the quantum dynamics simulation with the time-dependent Hamiltonian from approximating time-ordered evolution operators to solving discretized dynamical equations within a finite-dimensional projected subspace.
Time dependence is absorbed into a spectral expansion, reducing the simulation task to coherent inversion of a structured linear system while retaining high-precision state reconstruction.
In this sense, our method establishes a structural bridge between variational projection techniques and quantum linear-system algorithms, offering an alternative equation-driven route to time-dependent quantum dynamics.

The present method applies, in principle, to time-dependent Hamiltonians that admit an LCU representation with smoothly varying coefficients.
It does not rely on particular symmetries, conserved quantities, or commutation structures beyond this representation and coefficient smoothness, and is therefore applicable to general non-commuting dynamics.
This setting includes a broad class of problems in time-dependent quantum chemistry and atomic physics, where second-quantized Hamiltonians with smooth parameter dependence arise naturally.
Its practical effectiveness, however, depends on the existence of a compact projected subspace capable of capturing the relevant evolution.
For single-particle or few-channel systems such as ion-atom collisions, field-driven molecular processes, or driven few-level dynamics, the accessible subspace is typically much smaller than the full Hilbert space and dimensional reduction is physically well motivated.
In contrast, for strongly correlated many-body systems exhibiting extensive entanglement growth, the required ansatz dimension may scale unfavorably and the advantage of variational compression can diminish.
Developing systematic strategies for constructing compact and dynamically expressive variational ansatze for complex systems remains an open challenge.

Several directions for future work may further enhance the efficiency and scope of the present method.
On the algorithmic side, conditioning strategies that reduce the condition number of the global linear system would improve the scalability of the spectral formulation.
On the application side, further studies across a broader range of time-dependent systems would help delineate the practical scope of our method in increasingly complex dynamical settings.
The modular structure of the sequential formulation is naturally compatible with error mitigation techniques and could serve as a starting point for near-term experimental demonstrations on early fault-tolerant quantum processors.
Finally, combining our method with adaptive or problem-tailored ansatz constructions could extend the range of systems amenable to efficient variational compression.

In summary, this work reframes the dynamics simulation for a time-dependent Hamiltonian as a globally constrained algebraic problem rather than a sequence of operator exponentiations.
By embedding variationally projected dynamics within a coherent spectral linear-system formulation, it advances an equation-driven methodology that bridges time-dependent Hamiltonian simulation and quantum linear-system algorithms, expanding the conceptual landscape of fully quantum dynamical computation.


\begin{acknowledgments}
This work was supported by Innovation Program for Quantum Science and Technology with Grant No. 2021ZD0300200 and the National Natural Science Foundation of China with Grants No.~12374483 and No.~92365209.
\end{acknowledgments}






\nocite{*}
\bibliography{refs}







\end{document}